\documentclass[runningheads]{llncs}
\usepackage{graphicx}
\usepackage{amsmath}
\usepackage{amssymb}
\usepackage{booktabs}  

\DeclareMathOperator*{\argmin}{argmin} 
\begin{document}
%
\title{MRI Image Reconstruction via Learning Optimization Using Neural ODEs}
\titlerunning{MRI Image Reconstruction via Learning Optimization Using Neural ODEs}
%

\author{Eric Z. Chen  \and 
Terrence Chen  \and 
Shanhui Sun  
}  

\authorrunning{E. Chen et al.}
%
\institute{United Imaging Intelligence \\
Cambridge, MA 02140 \\
\email{\{zhang.chen, terrence.chen, shanhui.sun\}@united-imaging.com}\\
}

\maketitle              
\begin{abstract}
We propose to formulate MRI image reconstruction as an optimization problem and model the optimization trajectory as a dynamic process using ordinary differential equations (ODEs). We model the dynamics in ODE with a neural network and solve the desired ODE with the off-the-shelf (fixed) solver to obtain reconstructed images. We extend this model and incorporate the knowledge of off-the-shelf ODE solvers into the network design (learned solvers). We investigate several models based on three ODE solvers and compare models with fixed solvers and learned solvers. Our models achieve better reconstruction results and are more parameter efficient than other popular methods such as UNet and cascaded CNN. We introduce a new way of tackling the MRI reconstruction problem by modeling the continuous optimization dynamics using neural ODEs.

\end{abstract}

\section{Introduction}
One major limitation of Magnetic Resonance Imaging (MRI) is the slow data acquisition process due to the hardware constraint. To accelerate data acquisition, the undersampled k-space is often acquired, which causes aliasing artifacts in the image domain. Reconstruction of high-quality images from the undersampled k-space data is crucial in the clinical application of MRI. 

MRI image reconstruction is an inverse problem, in which the undersampling leads to information loss in the forward model and directly recovering the fully sampled image from undersampled data is intractable.  Compressed sensing (CS) provides the theoretical foundation for solving inverse problems by assuming the reconstructed image is sparse itself or in certain transformed domains. With the ability to learn complex distributions from data, deep learning has been applied to MRI reconstruction to learn optimal sparse transformations in an adaptive way  \cite{knoll2019deep,liang2019deep,ravishankar2019image}. Methods such as SToRM \cite{poddar2015dynamic} GANCS \cite{mardani2017deep}, and DAGAN \cite{yang2017dagan} follow such strategy and learn the prior distribution of the image from training data. On the other hand, several studies propose to tackle the inverse problem by learning the direct mapping from undersampled data to fully sampled data in the image domain \cite{lee2018deep}, k-space domain \cite{han2019k,akccakaya2019scan}, or cross domains \cite{zhu2018image}. Recent works extend this idea by learning such mapping in an iterative way using cascaded networks \cite{schlemper2017deep,huang2019mri,souza2019hybrid,souza2019dual,eo2018kiki,wang2018dimension,bao2019undersampled,aggarwal2018modl,gilton2019neumann}, convolutional RNN \cite{qin2018convolutional,wang2019pyramid} or invertible recurrent models \cite{putzky2019invert}. Many studies design the networks based on the iterative optimization algorithms used in CS \cite{sun2016deep,duan2019vs,hammernik2018learning,zhang2018ista,chen2019model}.

Ordinary differential equations (ODEs) are usually used to describe how a system change over time. Solving ODEs involves integration, most of which often have no analytic solutions. Therefore numerical methods are commonly utilized to solve ODEs. For example, the Euler method, which is a first-order method, is one of the basic numerical solvers for ODEs with a given initial value.  Runge–Kutta (RK) methods, which is a family of higher-order ODE solvers, are more accurate and routinely used in practice. 

Neural ODE \cite{chen2018neural} was introduced to model the continuous dynamics of hidden states by neural networks and optimize such models as solving ODEs.
With the adjoint sensitivity method, the neural ODE models can be optimized without backpropagating through the operations of ODE solvers and thus the memory consumption does not depend on the network depth \cite{chen2018neural}. ANODE \cite{gholami2019anode,zhang2019anodev2} further improves the training stability of neural ODEs by using Discretize-Then-Optimize (DTO) differentiation methods. 
\cite{yazdanpanah2019ode} applies ANODE to MRI reconstruction, where the residual blocks in ResNet \cite{he2016deep} are replaced with ODE layers. Neural ODE based methods are also applied to image classification \cite{paoletti2019neural} and image super-resolution \cite{he2019ode}. 

In this paper, we propose to formulate MRI image reconstruction as an optimization problem and model the optimization trajectory as a dynamic process using ODEs. We model the dynamics in the ODE with a neural network. The reconstructed image can be obtained by solving the ODE with off-the-shelf solvers (fixed solvers). Furthermore, borrowing the ideas from the currently available ODE solvers, we design network structures by incorporating the knowledge of ODE solvers. The network implicitly learns the coefficients and step sizes in the original solver formulation (learned solvers). We investigated several models based on three ODE solvers and compare neural ODE models with fixed solvers and learned solvers. 
We present a new direction for MRI reconstruction by modeling the continuous optimization dynamics with neural ODEs.


\section{Method}
MRI reconstruction is an inverse problem and the forward model is
\begin{align}
y = Ex + \epsilon,
\end{align}
where $x \in \mathbb{C}^M$ is the fully sampled image to be reconstructed,  $y \in \mathbb{C}^N$ is the observed undersampled k-space and $\epsilon$ is the noise. $E$ is the measurement operator that transforms the image into k-space with Fourier transform and undersampling. 
Since the inverse process is ill-posed, the following regularized objective function is often used:
\begin{align}
\argmin_x \frac{1}{2}||y-Ex||_2^2 + R(x),
\label{eq:1}
\end{align}
where $||y-Ex||_2^2$ is data fidelity term and $R(x)$ is the regularization term .
Eq.\ref{eq:1} can be optimized with gradient descent based algorithms, 
\begin{align}
x^{(n+1)} =  x^n - \eta [E^T (Ex^n-y) + \nabla R(x^n)], ~~ \text{for $n=1,\ldots,N$},
\label{eq:2}
\end{align}
where $\eta$ is the learning rate.

\subsection{ReconODE: Neural ODE for MRI reconstruction}

\begin{figure}[tbp]
\begin{center}
\includegraphics[width=\textwidth]{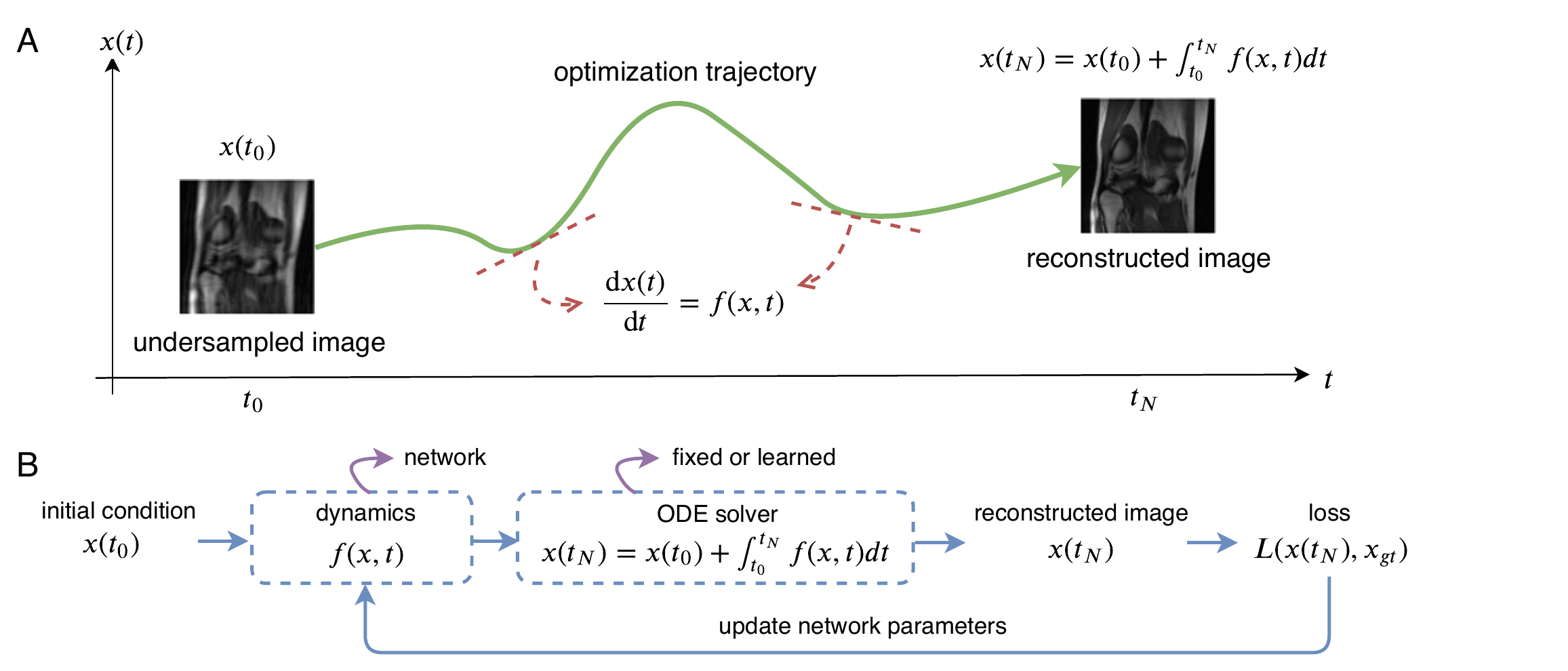}
\end{center}
\caption{An illustration of MRI reconstruction via modeling the optimization dynamics using neural ordinary differential equations. (A) MRI reconstruction is formulated as an optimization problem. The dynamic optimization trajectory is described by an ordinary differential equation. 
(B) We can model the dynamics with a neural network and perform image reconstruction as solving the ODE with the off-the-shelf solver. Alternatively, we can replace both the dynamics and the ODE solver with neural networks to implicitly learn the solver and perform image reconstruction.
} 
\label{fig:ODE_recon}
\end{figure}

The iterative optimization algorithm in Eq.\ref{eq:2} can be rewritten as
\begin{align}
x^{(n+1)} -  x^n &=  f(x^n,y,\theta), ~~\text{for $n=1,\ldots,N$}.
\label{eq:3}
\end{align}
The left hand side of Eq.\ref{eq:3} is the change of the reconstructed image between two adjacent optimization iterations. This equation essentially describes how the reconstructed image changes during the $N$ optimization iterations. The right hand side of Eq.\ref{eq:3} specifies this change by the function $f$. However, this change is described in discrete states defined by the number of iterations $N$. If we consider the optimization process as a continuous flow in time, it can be formulated as
\begin{align}
\dfrac{\mathrm{d}x(t)}{\mathrm{d}t} &=  f(x(t),t,y,\theta). 
\label{eq:ode}
\end{align}
Eq.\ref{eq:ode} is an ordinary differential equation, which describes the dynamic optimization trajectory (Fig. \ref{fig:ODE_recon}A). 
MRI reconstruction can then be regarded as an initial value problem in ODEs, where the dynamics $f$ can be represented by a neural network. 
The initial condition is the undersampled image and the final condition is the fully sampled image.  During model training, given the undersampled image and fully sampled image, the function $f$ is learned from data (Fig. \ref{fig:ODE_recon}B). During inference, given the  undersampled image as the initial condition at $t_0$ and the estimated function $f$, the fully sampled image can be predicted by evaluating the ODE at the last time point $t_N$,
\begin{align}
x(t_N) = x(t_0) + \int_{t_0}^{t_N} f(x(t),t) dt, 
\label{eq:integral}
\end{align}
where $y$ and $\theta$ are omitted for brevity. An arbitrary time interval [0,1] is set as in \cite{chen2018neural}. Evaluating Eq.\ref{eq:integral} involves solving the integral, which has no analytic solution due to the complex form of $f$. The integral needs to be solved by numerical ODE solvers. 

Next, we will introduce two models that either use the off-the-shelf solvers (fixed solvers) or learn the solvers by neural networks (learned solvers).

\subsection{ReconODE with fixed solvers}
The Euler solver is a first-order ODE solver, which employs the discretization method to approximate the integral with a step size $h$ in an iterative way,
\begin{align}
x_{n+1} &= x_{n} + hf(x_{n}, t_n),
\label{eq:euler_solver}
\end{align}
which is often called the step function of a ODE solver.
More complicated methods such as RK solvers with higher orders can provide better accuracy. The step function for a general version of RK methods with stage $s$ is, 
\begin{align}
x_{n+1} &= x_{n} + \sum_{i=1}^{s}a_i F_i\\
F_1 &= hf(x_{n}, t_n)\\
F_i &= hf(x_{n}+\sum_{j=1}^{i-1}b_{ij} F_j, t_n+c_i h), ~~ \text{for $i=2,\ldots,s$},
\label{eq:general_RK4}
\end{align}
where $a_i$, $b_{ij}$ and $c_i$ are pre-specified coefficients. 
As an example, for RK2 and RK4 solvers, the coefficients are specified as $s=2, a_1=a_2=\frac{1}{2}, b_{21}=c_2=1$ and $s=4, a_1=a_4=\frac{1}{6}, a_2=a_3=\frac{2}{6}, b_{21}=b_{32}=c_2=c_3=\frac{1}{2},b_{43}=c_4=1 $, respectively (all other coefficients are zeros if not specified). 

By using one of off-the-shelf ODE solvers to evaluate Eq.\ref{eq:integral} numerically, the MRI image can be reconstructed as solving a neural ODE. In our experiment, we model the dynamics $f$ as a CNN (Fig. \ref{fig:RKNet}B) with time-dependent convolution (Fig. \ref{fig:RKNet}A), in which the time information was incorporated into each convolutional layer by concatenating the scaled time with the input feature maps \cite{chen2018neural}. To train such models, we can either backpropagate through the operations of solvers (ReconODE-FT) or avoid it with the adjoint sensitivity method \cite{chen2018neural,gholami2019anode} to compute the gradient (ReconODE-FA). ``F" stands for fixed solvers,  ``T" and ``A" indicate backpropagating through solvers and using the adjoint method, respectively. A data consistency layer \cite{schlemper2017deep} is added after the output from the ODE solver.

\subsection{ReconODE with learned solvers}
We now extend the above idea further: 
instead of using the known ODE solvers directly, we incorporate the knowledge of the solvers into the network and let the network learn the coefficients and step size. We can write the step function in Eq.8-10 as
\begin{align}
x_{n+1} &= x_{n} + G(F_1,...,F_s; \omega)\\
F_1 &= G_1(x_{n}, t_n, y; \theta_1)\\
F_i &= G_i(x_{n},F_1,...,F_{i-1},t_n,y; \theta_i), ~~ \text{for $i=2,\ldots,s$},
\label{eq:network_RK}
\end{align}
where $G$ and $G_i$ are neural networks with parameters $\omega$ and $\theta_i$. The basic building block  $G_i$ is a CNN with five time dependent convolutional layers, which not only learns the dynamics $f$ but also the coefficients and step size in the original solver. 
Furthermore, since we observe that during optimization more high-frequency details are recovered and in the original solver, $F_{i+1}$ is evaluated ahead of $F_i$ in time, we expect $F_{i+1}$ to recover more details. Thus to mimic such behavior, the network $G_{i+1}$ has a smaller dilation factor than $G_{i}$ to enforce the network to learn more detailed information. Based on the knowledge that $a_i \in (0,1)$ in $\sum_{i=1}^{s}a_iF_i$, we replace the weighted sum in Eq.8 with an attention module $G$.

\begin{figure}[!htb]
\begin{center}
\includegraphics[width=\textwidth]{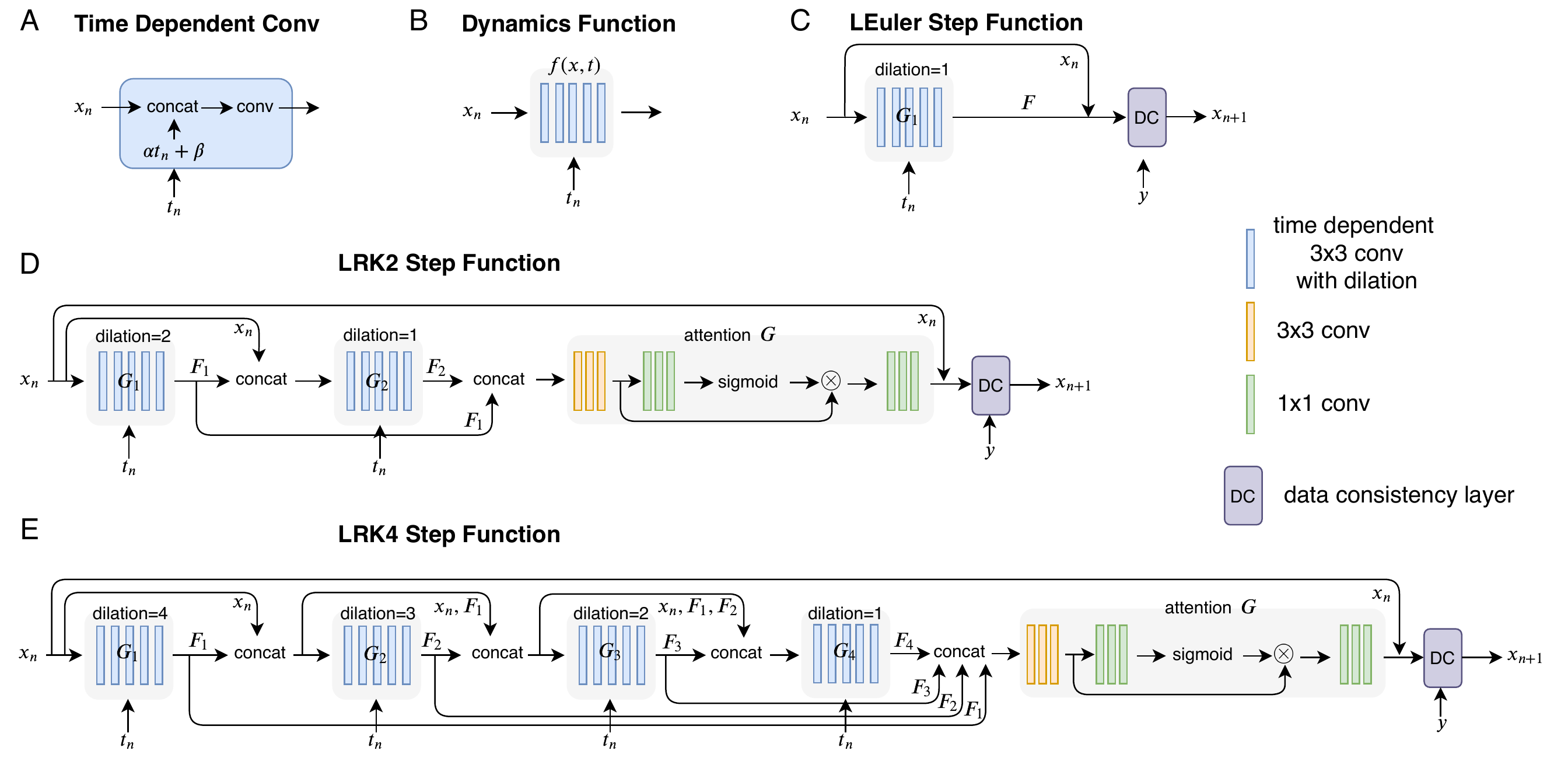}
\end{center}
\caption{The building blocks and overall structure of proposed ReconODE networks. (A) In the time-dependent convolutional layer, the time after a linear transformation and input feature maps are concatenated and then fed into the convolution operation. (B) The dynamics in ODE is represented by a CNN with five time-dependent convolutional layers. Using off-the-shelf solvers, the ODE can be solved to obtain reconstructed images. (C-E) We extend this idea and propose three neural networks to learn step functions of Euler, RK2 and RK4 solvers based on the original solver formulations. Different dilation factors are utilized to learn multi-scale features. The ReconODE-LT network is a cascade of the corresponding solver step functions. The parameters are shared across cascade iterations.}
\label{fig:RKNet}
\end{figure}

We design three networks based on the step functions of Euler, RK2 and RK4, respectively (Fig. \ref{fig:RKNet}C-E), named as ReconODE-LT (``L" indicates learned solvers and ``T" is backpropagation through solvers). The final network is the cascade of solver step functions and the parameters are shared across iterations. 
 
One potential drawback of incorporating the ODE solver into the network is the increased GPU memory usage during training, especially for complicated ODE solvers. To alleviate this problem, we adopt the gradient checkpoint technique \cite{chen2016training} to dramatically reduce the memory consumption while only adding little computation time during training, where the intermediate activations are not saved in the forward pass but re-computed during the backward pass.  

\subsection{Model training and evaluation}
We used the single-coil fastMRI knee data \cite{zbontar2018fastmri}. There are 34,742 2D slices in the training data and 7,135 slices in the validation data. The fully sampled k-space data were undersampled with acceleration factors (AF) 4 and 8, respectively. We compared our models with UNet \cite{ronneberger2015u,zbontar2018fastmri} and modified it for data with real and imaginary channels and added a data consistency layer \cite{schlemper2017deep}. The cascade CNN model (D5C5) \cite{schlemper2017deep} and KIKI-Net \cite{eo2018kiki} were also included for comparison. For ReconODE-FA models, we adapted the code from \cite{gholami2019anode}.
For a fair comparison, we applied the default settings of channel=32 from fastMRI UNet and N=5 from cascade CNN to all models if applicable.
All models were trained using $loss = L1+0.5 \times SSIM$ and RAdam  \cite{liu2019variance} with Lookahead \cite{zhang2019lookahead}.  Results were evaluated on the fastMRI validation data using PSNR and SSIM \cite{zbontar2018fastmri}. 

\section{Result}
\begin{figure}[!htb]
\begin{center}
\includegraphics[trim=0 0 0 0,clip, width=\textwidth]{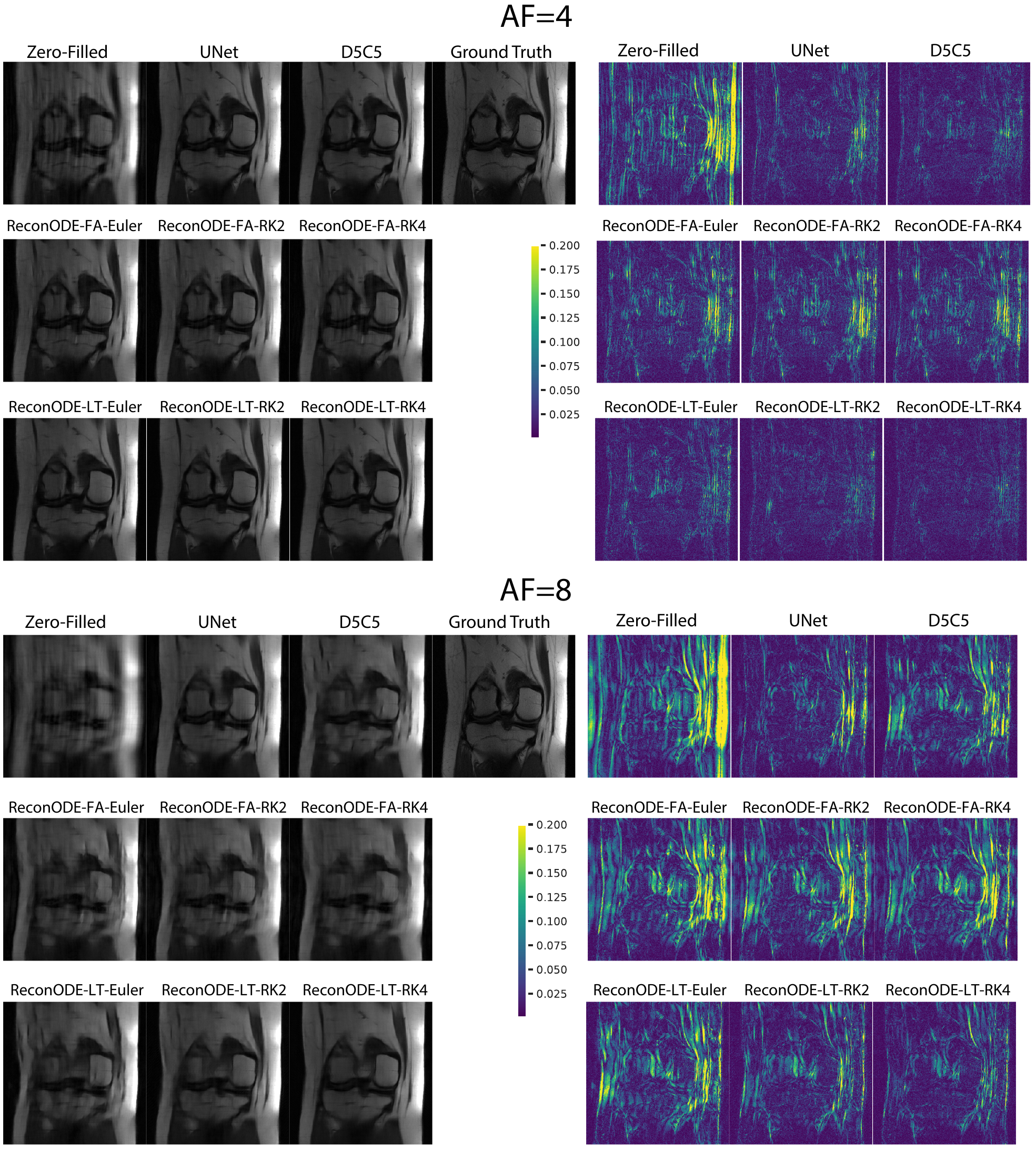}
\end{center}
\caption{Examples of reconstructed images and corresponding error maps. Results of other models are omitted due to the space limit.} 
\label{fig:recon_result}
\end{figure}

Table \ref{table:validation} shows the reconstruction results on the fastMRI validation dataset. UNet, which learns the direct mapping between the undersampled image to the fully sampled image, has the largest number of parameters and has about 20 to 100 times more parameters than ReconODE models. D5C5 that learns the mapping in an iterative way has slightly better SSIM in 4X but worse SSIM in 8X than UNet. 
ReconODE-LT-Euler with only $0.9\%$ parameters of UNet achieves similar results as UNet at 4X. ReconODE-LT-RK2 with $64\%$ parameters of D5C5 has similar results as D5C5 at 4X and 8X. The ReconODE-LT-RK4 achieves the best PSNR and SSIM at both 4X and 8X among all models. 
Fig. \ref{fig:recon_result} shows examples of reconstructed images. ReconODE-LT-RK4 achieves the smallest error among all models, which is consistent with the quantitative results. 

Using more sophisticated but fixed ODE solvers in ReconODE-FT as well as ReconODE-FA models does not seem to significantly improve the reconstruction results. This is in line with previous results \cite{gholami2019anode}.
However, with more complicated but learned ODE solvers, the performance of ReconODE-LT models is improved. Also, ReconODE-LT models outperform ReconODE-FT models with the same ODE solver. These results indicate that learning the ODE solver by the network is beneficial. To demonstrate the effectiveness of the attention in our network, we also trained a ReconODE-FT-RK4 model without attention. We observe that the SSIM at 4X drops from 0.733 (with attention) to 0.726 (without attention). 
ReconODE-FT models have overall better performance than ReconODE-FA models, which suggests that backpropagation through solvers may be more accurate than the adjoint method.
All the improvements described above are statistically significant ($p<10^{-5}$).

With the help of the gradient checkpoint technique,
the GPU memory usage for ReconODE-LT-RK4 can be significantly reduced ($73\%$) with only 0.13 seconds time increase during training.
We did not observe any difference in testing when applying the gradient checkpoint (Table \ref{table:benchmark}).

We initially tested the original neural ODE model \cite{chen2018neural} but the performance was poor (4X SSIM 0.687) and the training was very slow, which may be due to the stability issue \cite{gholami2019anode,zhang2019anodev2}. 

\begin{table}[!htb] 
\caption{Quantitative results on the fastMRI validation dataset. 
}
\centering
  \begin{tabular}{lccccc}
    \toprule
     & & \multicolumn{2}{c}{PSNR}  & \multicolumn{2}{c}{SSIM}\\
    \cmidrule(r){3-4} \cmidrule(r){5-6}
    \multicolumn{1}{c}{Model} & Param &  4X     & 8X  & 4X     & 8X \\
    \midrule
    UNet & 3.35M &    31.84 & 29.87 & 0.7177 & 0.6514 \\ 
    KIKI-Net &  0.30M &  31.83 & 29.38 & 0.7178 & 0.6417  \\
    D5C5 &  0.14M &  32.11 & 29.45 & 0.7252 & 0.6425  \\
    \midrule
    ReconODE-FA-Euler   & 0.03M &     31.21 & 28.48  & 0.7035 & 0.6172 \\ 
    ReconODE-FA-RK2  & 0.03M &     31.25 & 28.50 & 0.7045 & 0.6175 \\ 
    ReconODE-FA-RK4   & 0.03M &     31.22 & 28.57 & 0.7027 & 0.6183 \\ 
    \midrule
    ReconODE-FT-Euler & 0.03M &   31.83   & 29.09 & 0.7188  &  0.6355  \\
    ReconODE-FT-RK2  &  0.03M &   31.77 &  29.04  &   0.7193 &   0.6345 \\
    ReconODE-FT-RK4  &  0.03M &  31.82 & 29.16  &  0.7204 & 0.6388 \\ 
    \midrule
    ReconODE-LT-Euler  & 0.03M &     31.86 & 29.11 & 0.7194 & 0.6363   \\ 
    ReconODE-LT-RK2  &  0.09M &   32.19 &  29.77  &  0.7292 & 0.6498   \\
     ReconODE-LT-RK4  &  0.15M &   \textbf{32.39} &  \textbf{30.27}   &  \textbf{0.7333} &   \textbf{0.6617}\\
    \bottomrule
  \end{tabular}
  \label{table:validation}
\end{table}

\begin{table}[!htb]
  \caption{Benchmark the ReconODE-LT-RK4 model with or without gradient checkpoint technique. 
  }
  \centering
  \begin{tabular}{cccccc}
    \toprule
     &&\multicolumn{2}{c}{Train}  & \multicolumn{2}{c}{Test}\\
    \cmidrule(r){3-4} \cmidrule(r){5-6}
    \multicolumn{1}{c}{Gradient Checkpoint}   & Param &  GPU     & Time  & GPU     & Time \\
    \midrule
    Y   & 0.15M &   3.7G   & 1.01s & 1.4G &   0.15s  \\ 
    N   & 0.15M &   13.6G  & 0.88s & 1.4G &   0.15s  \\ 
    \bottomrule
  \end{tabular}
  \label{table:benchmark}
\end{table}

\section{Discussion and conclusion}
In this paper, we propose an innovative idea for MRI reconstruction. We model the continuous optimization process in MRI reconstruction via neural ODEs. Moreover, our proposed ReconODE-LT models integrate the knowledge of the ODE solvers into the network design. 

Since the fastMRI leaderboard\footnote{https://fastmri.org/leaderboards} was closed for submission, we only evaluated the results on the validation dataset. Our results on validation data are not directly comparable to the top leaderboard results on test data, which train on validation data with more epochs \cite{sriram2020end} and model ensembles \cite{hammernik2019sigma}. Moreover, compared to models such as E2E-VN (30M) \cite{sriram2020end}, PC-RNN (24M) \cite{wang2019pyramid}, and Adaptive-CS-Net (33M) \cite{pezzotti2019adaptive}, our ReconODE models are much smaller ($\leq0.15$M parameters). 
The proposed methods achieve comparable performance with much smaller model size \cite{putzky2019invert}. We expect further studies will lead to a better trade-off between performance and model size.
As we intend to propose a new framework rather than a restricted model implementation, further boost of performance can be expected by using larger networks for ODE dynamics and more complicated solvers.
This paper provides potential guidance for further research and extension using neural ODE for MRI reconstruction.

%
%
%
\bibliographystyle{splncs04}
\bibliography{ref}

\end{document}